\documentclass[aps,prl,showkeys,showpacs,10pt]{revtex4-1}
\usepackage{amssymb}
\usepackage{hyperref}
\usepackage{graphics,graphicx}
\usepackage{dcolumn}
\usepackage{amsbsy}
\usepackage{amsmath}
\usepackage{epsfig}
\usepackage{color}
\usepackage{textcomp}
\begin{document}
 \title{Tunneling conduction in graphene/(poly)vinyl alcohol composites}
\author{Sreemanta Mitra$^{1,2}$}
\email[]{sreemanta85@gmail.com}
\author{Sourish Banerjee$^{2}$}
\author{Dipankar Chakravorty$^{1,\dag}$}
\email[]{mlsdc@iacs.res.in}
\affiliation{
$^{1}$
 MLS Prof.of Physics' Unit,Indian Association for the Cultivation of Science, Kolkata-700032, India.\\ }
\affiliation{
$^{2}$
Department of Physics, University of Calcutta, Kolkata-700009, India.\\}

\begin{abstract}
Graphene/(Poly)vinyl alcohol (PVA) composite film with thickness $60 \mu m$ were synthesized by
solidification of a PVA solution comprising of dispersed graphene nanosheets. The close proximity of the graphene
sheets enables the fluctuation induced tunneling of electrons to occur from one sheet to another. The dielectric data show that the present system can be simulated to 
a parallel resistance-capacitor network. The high frequency exponent of the frequency variation of the ac conductivity
indicates that the charge carriers move in a two-dimensional space. The sample preparation technique will be helpful for synthesizing flexible conductors.

\end{abstract}
\maketitle
\section{Introduction}
Recently, graphene and graphene-based materials have attracted considerable attention not only
because of deriving rich physics from it, but also to exploit these for fabricating simpler and more 
efficient devices \cite{geimsc,novonat,stancovichcarbon,lotya,cnrjpcl,cnrjmc,cnracm,reina,castroneto}.
To design graphene-based nanoscale devices, quantum transport mechanism has to be kept in view.
Synthesis of composites with single or few layer graphene sheets has been reported recently. This requires large scale production of graphene, and their homogeneous
distribution in various matrices\cite{stankovichnat}. 
Several systems have been studied of which transparent and electrically conducting graphene/silica \cite{watchnl} and graphene/polystyrene composites made by complete exfoliation
of graphite\cite{stankovichnat} may be mentioned. Graphene/(poly)aniline composites\cite{almashat} have been used for hydrogen gas sensing.
 Theoretically it had been shown that graphene's
volume fraction, or the aspect ratio of the graphene sheet had a profound effect on the resistance of the graphene-based polymer composites \cite{apl95}.
We had previously synthesized graphene/PVA composite, with a particular concentration of graphene sheets
in the polymer matrix to study the magnetodielectric effect\cite{mitrajpcc}. In the latter, we have
shown that electrical conductivity arises due to hopping of electrons between the localized states
provided by the graphene sheets. From the above, it can be seen that so far electrical conductivity
involving graphene has been investigated  either in its percolative configuration\cite{krupkaapl} or
in a composite structure\cite{stankovichnat,watchnl}. In the present work, our objective is to study the conduction mechanism 
in a graphene based polymer composite, where the graphenes will be at  close proximity to each other but not in percolative configuration. 
The details of our findings are reported in this letter. 
 \section{Experimental}
A simple chemical method was employed to synthesize the graphene/(poly)-vinyl alcohol(PVA) composite. The synthesis of
graphene from the chemical exfoliation of graphite oxide (GO) has been described elsewhere \cite{stancovichcarbon,mitrajpcc}.
Firstly, a modified Hummers method was employed to synthesize graphene oxide from high purity graphite flakes
(LOBACHEMIE). 0.01 g of GO was dispersed in 10 mL water, and then 6 mL of hydrazine hydrate was mixed with it. Ammonia solution 
was added to keep the pH of the mixture at 10. After 3h of stirring graphene dispersion was achieved.   
\par
(Poly)vinyl alcohol (PVA 0.505 g) powder (as obtained from S-d fine-Chem,India,$M_{w}\approx14000$) was dissolved in 20 mL of water and stirred at 333 K
for 3 h. to form a homogeneous polymer solution. 6 mL of dispersed graphene was mixed with PVA (12 mL)solution and 
stirred for 4 h. to form a homogeneous mixture. To obtain the film, the mixture was cast on a teflon coated
petri-dish and solidified at room temperature. 
It is to be noted here that the 
 preparation entailed 1.95{\%} GO and 98.05{\%} of PVA
(by weight) as precursors.
The graphene was characterized previously by 
Fourier Transform Infrared Spectroscopy (FTIR) and transmission electron microscopy \cite{mitrajpcc}.
Electrical measurements were carried out after applying silver electrodes (silver paint supplied by M/S Acheson Colloiden,The Netherland)
on both sides of a piece of graphene-PVA composite film using a 
Keithley 617 electrometer. The current voltage characteristic was measured using a Kiethley 2400 source meter.
For dielectric measurements an Agilent 4980 precision LCR meter was used. 
\section{Results and Discussion} 
It was observed \cite{mitrajpcc} from the FTIR spectra that, hydroxyl and epoxide functional groups present on the basal plane of GO, 
and carbonyl and carboxyl groups at the edges of GO were removed (absence of any C=O) due to the chemical reaction 
of GO with hydrazine hydrate.\\
\begin{figure}
 \centering
 \includegraphics[width=8.5cm]{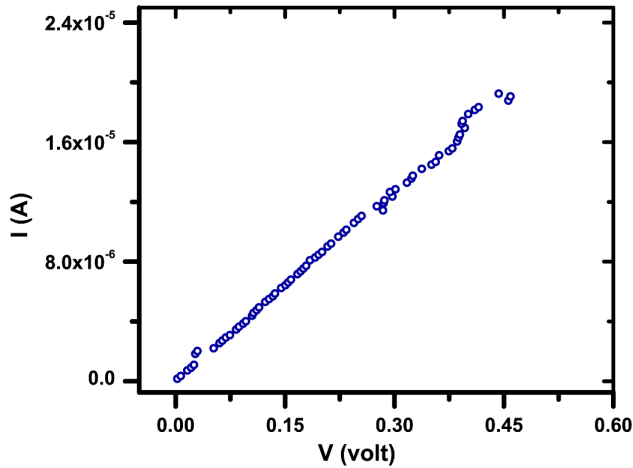}
 \caption{ Variation of current with voltage measured at room temperature (306 K).}
 \label{f1}
\end{figure}
A typical voltage current characteristic for the composite, obtained at room tempreratre is shown in figure \ref{f1}. The linear behaviour indicates a non-blocking
nature of the electrodes used. 

\begin{figure}
\centering
\includegraphics[width= 8.5cm]{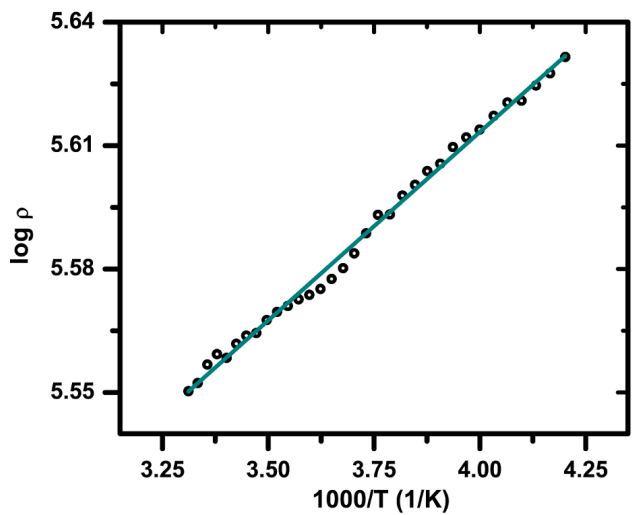}
\caption{ Variation of logarithm of resistivity with T${^{-1}}$ for the composite.The solid line represents the 
least squared fitted curve,with equation\ref{eq.1}.}
\label{f2}
   \end{figure}

The variation of logarithm of dc resistivity as a function of $T^{-1}$ for the composite film of thickness 0.006 cm
is shown in figure \ref{f2}. 
This semiconductor like behaviour in the temperature dependence of the resistivity can be due to the existence of
potential barriers between highly conducting regions. In this kind of situation the conduction arises due to 
hopping or tunneling of the charge carriers from one conducting island to another. 
We have fitted the experimental data with the fundamental equation of the activated process, 
\begin{equation}
 \rho=\rho_{o} \exp{(\frac{\phi}{k_{B}T})}                    
\label{eq.1}
\end{equation}
where, $\rho$ is the resistivity, $k_{B}$ is Boltzmann constant, $\phi$ is the activation energy and T is 
the absolute temperature. 

The straight line in the figure \ref{f2} shows the least square fitted curve to equation \ref{eq.1}. The points indicate the experimental data.
The slope of the straight line was extracted as $0.0805 \pm 0.001$,from which we have calculated the activation energy $\phi$ as 0.016 eV. The low value of activation energy, suggests that the charges tunnel from one graphene sheet to another through the polymer.
Given that each graphene is coated with the polymer, which acts as the potential barrier for the inter graphene 
hopping, it is likely that electrical conductivity in this system is
governed by tunneling between conductive regions. 
In order to get more insight about the conduction mechanism, we invoked the model of fluctuation induced tunneling 
of charge carriers between highly conducting regions for inhomogeneous conductors \cite{pingprb}. This model for disordered material is mainly used for the conduction of charge carriers from large sized conductors, separated by small insulating regions. The concentration of the graphene sheets dispersed 
in PVA matrix,being much higher than that used earlier\cite{mitrajpcc}, the separation between graphene sheets is very small, hence this type of transport mechanism is to be expected. 
The temperature variation of the conductivity in this fluctuation induced tunneling model, is given by, \cite{pingprb},

\begin{equation}
 \sigma_{dc}=\sigma_{0} \exp{(- \frac{T_{1}}{T+T_{0}})}
\label{eq.2}
\end{equation}
\begin{figure}
\centering
\includegraphics[width= 8.5cm]{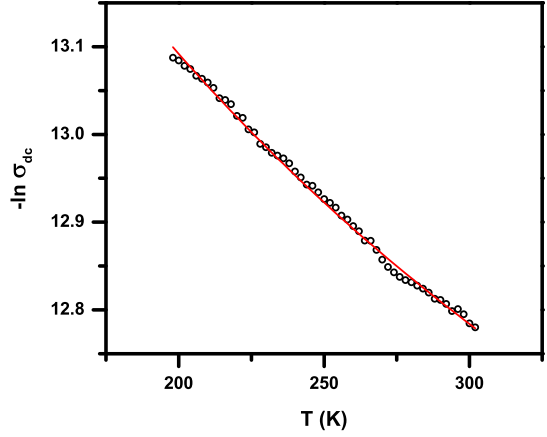}
\caption{ Variation of logarithm of dc conductivity with temperature for the composite.The solid line represents the  least squared fitted curve,with equation\ref{eq.2}.}
\label{f3}
   \end{figure}
where, $T_{1}$ is the temperature required for an electron to cross the insulator gap between conductive regions, and  $T_{0}$ is the temperature above which thermal activated conduction over
potential barrier begins to occur. The experimental data and the least square fitted curve with equation \ref{eq.2} are shown in fig.\ref{f3}. 
$T_{1}$ and $T_{0}$ are material constants and related to the energy barrier U by, 
\begin{equation}
T_{1}=\frac{16AU^{2}}{8\pi k_{B}we^{2}}
\label{eq.3}
\end{equation}
and 
\begin{equation}
T_{0}=\frac{2hT_{1}}{\pi w (2mU)^{\frac{1}{2}}}
\label{eq.4}
\end{equation}
 where, e and m are the electronic charge and mass respectively, w the intergraphene gap width, and A
the area of capacitance formed at the junction, $k_{B}$ stands for the Boltzman Constant. Our fitting of the 
experimental data with eq.\ref{eq.2} yielded the values of $T_{1}$ and $T_{0}$, which were used as parameters,
as 746.23 K (or 0.06eV) and 244.8 K (or 0.021eV) respectively. Since we do not have the data to observe the distribution 
of $T_{1}$ and $T_{0}$, these observed values, should be interpreted as indicative of the median values of these parameters.   
However, it is worth noting, that the activation energy is smaller than that  obtained previously for granular metal
films \cite{webbjap}. Previously,\cite{mitrajpcc} we have reported that for lower concentration of the graphene
sheets in the PVA matrix, the activation energy has a value of 0.38 eV. 
In this particular work we have stuck to fluctuation induced tunneling as the conduction mechanism, and ruled out the possibility of other activated processes,
such as, Frenkel-Poole and Fowler-Nordheim tunneling or hopping mechanisms which are responsible for charge conduction in this type of disordered materials. The reasons for this are as follows.
The experimental data are significantly different from the 
\begin{equation}
 ln \rho \sim [\frac{E^{\frac{1}{2}}}{T}] 
\end{equation}
 of Frenkel-Poole type or
\begin{equation}
 ln (\frac{j}{E^{2}}) \sim [\frac{1}{E}]
\end{equation}
 of Fowler-Nordheim type of tunneling. The hopping mechanism were
excluded since the resistivity data doesnot follow the stretched exponential function of the type $\exp [a/T^{\alpha}]$ 
for a single value of $\alpha$\cite{mott}.  

 \begin{figure}
\centering
\includegraphics[width= 8.5cm]{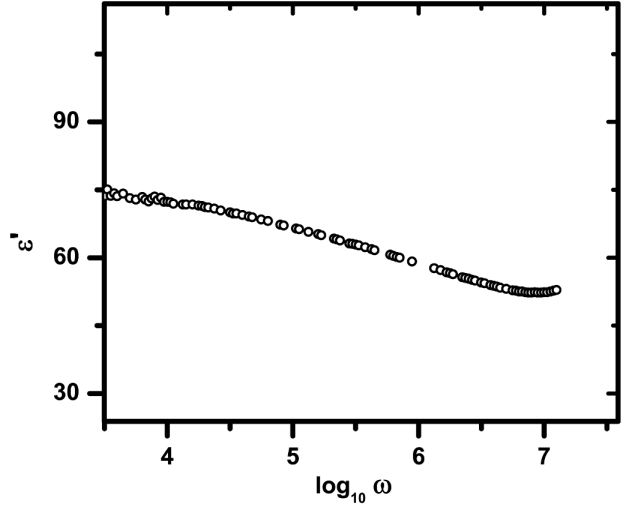}
\caption{Variation of real part of the dielectric permittivity ${(\epsilon{'})}$ of the 
nanocomposite as a function of the frequency measured at 306 K .}
\label{f4}
   \end{figure}
 \begin{figure}
\centering
\includegraphics[width= 8.5cm]{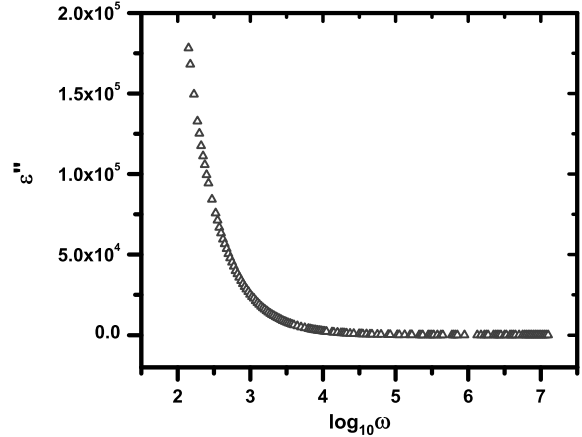}
\caption{Variation of imaginary part of the dielectric permittivity ${(\epsilon{''})}$ of the 
   composite with the frequency measured at 306 K.}
\label{f5}
   \end{figure}
 \begin{figure}
\centering
\includegraphics[width= 8.5cm]{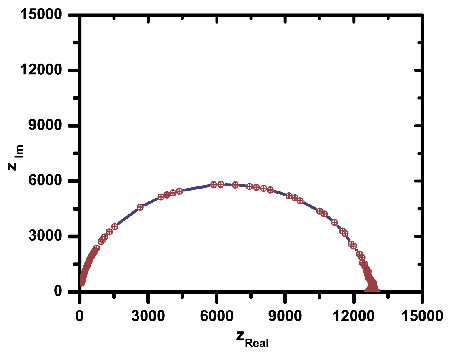}
\caption{Cole-Cole diagram of the composite measured at 306 K.}
\label{f6}
   \end{figure}
The dielectric permittivity of the composite film was measured and we have delineated the real and 
imaginary parts of the same. The variation of real ${(\epsilon{'})}$ and imaginary ${(\epsilon{''})}$ parts of 
the dielectric permittivity with frequency measured at room temperature (306 K) are shown in figures \ref{f4} and \ref{f5} respectively. It is seen that, real part of the dielectric permittivity ${(\epsilon{'})}$ decreases slowly as a function of frequency in the range studied,
 whereas the imaginary part ${(\epsilon{''})}$
decreases drastically from a high value. This type of variation is expected in case of a circuit consisting of a 
parallel combination of resistance and capacitance. We believe that the graphene films form the resistive
network whereas, the PVA film makes up the capacitive element in the network system \cite{hippel}.
The high value of the imaginary part of dielectric permittivity also signifies that there 
has been a high loss factor associated with the sample for a resistive network. 
A Cole-Cole diagram of the composite has been shown in figure \ref{f6}. A single semi-circle signifies that
there is one conduction mechanism operative in the composite. From the Cole-Cole diagram,
the dc resistance was extracted from $z_{im} $ $ \rightarrow 0 $ 
and found to be around 13 k$\varOmega$. The linear fitting of the 
current voltage data[figure\ref{f1}] shows the resistance of the composite to be 23.6 $ k\varOmega$ which is of the same order of magnitude as that obtained from the Cole Cole diagram.
 \begin{figure}
\centering
\includegraphics[width= 8.5cm]{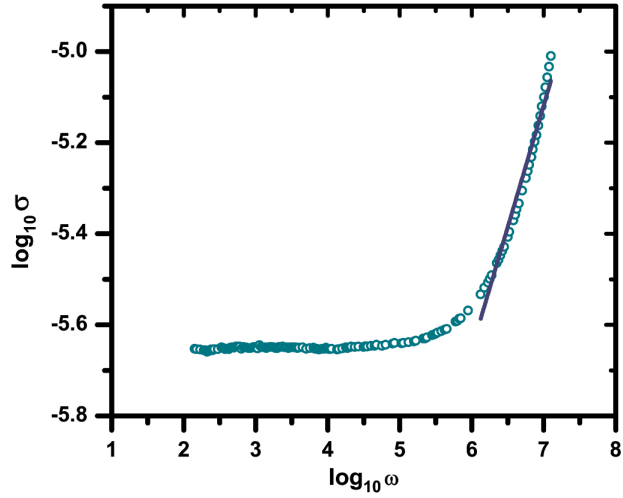}
\caption{Variation of logarithm of ac conductivity with logarithm of frequency measured at 306 K.The solid line 
represents the least squared fitted curve. }
\label{f7}
   \end{figure} 
   The frequency dependent dispersion of conductivity is one of the characteristic features of electronic conduction in the disordered material\cite{dyrermp}. The frequency independent part at low frequency is observed to show a 
 freuency dependence after a particular characteristic frequency, $\omega_{0}$. After  $\omega_{0}$ the conductivity 
 increases in a power law manner, 
\begin{equation}
 \sigma=A \omega^{\beta}
\label{eq.5}
\end{equation}
The variation of logarithm of ac conductivity as a function of logarithm of angular frequency
for the graphene/PVA composite measured at 306 K is shown in figure \ref{f7}. 


We have fitted the conductivity data in the high frequency range, with the above mentioned equation, where $\sigma$ is the ac conductivity
\textquoteleft A \textquoteright is a constant, $\omega$ is the frequency of the applied electric field and $\beta$ is the exponent. From the fitting
the value of the exponent ($\beta$) is determined.
The value of \textquoteleft $\beta$ \textquoteright determines the dimension in which the conduction is
taking place \cite{funkessc,sidebottomprl,donaldjcp1,donaldjcp2}.
For one dimensional conduction the value of \textquoteleft$ \beta $\textquoteright lies around 0.3,whereas,for two 
and three dimensions its value is around 0.5 and 0.66 respectively \cite{funkessc,sidebottomprl,donaldjcp1,donaldjcp2}.
The value of  \textquoteleft $\beta$ \textquoteright in this case is extracted as $0.51 \pm 0.01$, which confirms the electronic motion to be operative in two-dimensions. This is due to the fact that the conduction is governed by the charge carriers of a two dimensional path formed by graphene sheets in close proximity to each other. 

 \begin{figure}
\centering
\includegraphics[width=15cm]{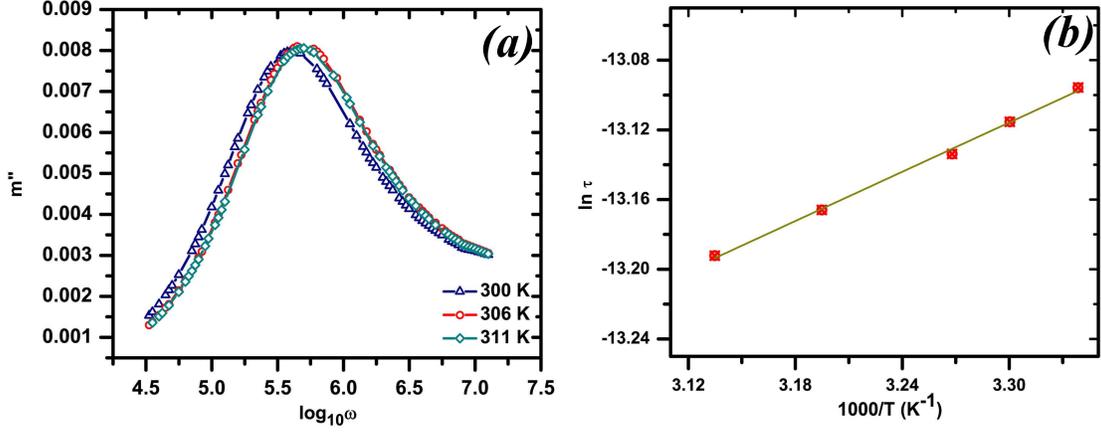}
\caption{(a) Variation of dielectric modulus with logarithm of frequency measured at different temperatures. (b)
The variation of logarithm of relaxation time with inverse of temperature. The straight line fitting gives the value of the activation 
energy for the relaxation process.}
\label{f8}
   \end{figure} 
Figure \ref{f8}(a) shows the variation of the imaginary part of dielectric modulus with angular frequency at different temperatures.

Dielectric modulus $(M^{*})$ is defined as \cite{hodgejncs} 
\begin{equation}
 M^{*}=\frac{1}{\epsilon^{*}}
\label{eq.6}
\end{equation}
 where, $\epsilon^{*}$ is the dielectric permittivity. The curves show prominent peaks,whose position is seen to shift 
towards the higher frequency side,when the temperature is increased. From the peak positions we have delineated the 
values of the relaxation time $(\tau)$ for the composite using the relation; 
\begin{equation}
 \omega \tau=1
\label{eq.7}
\end{equation}
In  figure \ref{f8}(b), the variation of $ln \tau$ with $T^{-1}$ has been shown. From the slope of this Arrhenius plot the value of the activation energy for the relaxation process was calculated  by employing,
\begin{equation}
 \tau=\tau_{0}exp(\frac{E_{a}}{k_{B}T})
\label{eq.8}
\end{equation}
where $E_{a}$ is the activation energy for the relaxation mechanism. The slope of the straight line comes out to be 
$0.471 \pm 0.013$ and the corresponding $E_{a}$ as  0.04 eV. This is of the same order of magnitude as the activation 
energy estimated from the dc resistivity data. This can be explained as follows.
The imaginary part of dielectric permittivity ${(\epsilon{''})}$ can be written as 

\begin{equation}
 \epsilon{''}=\epsilon{'}tan\delta
\label{eq.9}
\end{equation}
 
where $\epsilon{'}$ is the real part of the dielectric permittivity and $tan\delta$ is the dissipation factor.
The present sample system has been described earlier as a parallel combination of resistance and capacitance,
the former contributed by the two-dimensional conductivity of graphene/PVA composite and the latter arising
out of the PVA phase. From eq.(6) therefore, we get,
\begin{equation}
 \epsilon{''}=\frac{\epsilon{'}}{\omega\rho\epsilon{'}}=\frac{1}{\omega\rho}
\label{eq.10}
\end{equation}
where, $\omega$ is the angular frequency and $\rho$ is the resistivity of the graphene/PVA composite. 
Hence, $\epsilon{''}$ should have a temperature variation identical to that of $\rho$. 
 
\section{Conclusion}
 In summary, Graphene/PVA composite has been synthesized such that graphene sheets in two dimensions
form a network with the sheets in close proximity to each other. The electrical resistivity variation
with temperature shows a low activation energy $\sim 0.02 eV$ indicating an electron tunneling process to be operative. We observed that the fluctuation induced tunneling between graphene sheets was the 
mode of electronic conduction in the composite.
 The dc I–V characteristic was ohmic,
while in general the ac conductivities displayed two regions:
a low frequency region of constant conductivity and a high
frequency region with conductivity increases as a power law.
High frequency exponent extracted from the ac conductivity data confirms the charge carrier
movement to be taking place in a two-dimensional space. Dielectric modulus data at different 
temperatures were analyzed in terms of a relaxation mechanism, which gave an activation energy
similar to that found from dc resistivity data. This is shown to be consistent with the structural aspects
of the composite studied here. Using the present fabrication technique, it should be possible to prepare flexible 
 conductors with different conductivities. 
\acknowledgments
 Support for this work was derived from an Indo-Australian Project on nanocomposites for Clean Energy granted
 by Department of Science and Technology, Govt. of India,
 New Delhi. SM thanks University Grants Comission,New Delhi,for awarding Senior Research Fellowship. 
 DC thanks Indian National Science Academy, New Delhi, for an Honorary Scientist's position.

%
\end{document}